# Deep Learning-based Intraoperative MRI Reconstruction

**Jon André Ottesen**[1, 2, 3*], **Tryggve Storas**[3], **Svein Are Sirirud Vatnehol**[4, 5, 6], **Grethe Løvland**[4], **Einar O. Vik-Mo**[10], **Till Schellhorn**[1], **Karoline Skogen**[1], **Christopher Larsson**[1, 7], **Atle Bjørnerud**[1, 2, 3], **Inge Rasmus Groote-Eindbaas**[1,8], **Matthan W.A. Caan**[1,9]

[1]Computational Radiology & Artificial Intelligence (CRAI) Research Group, Division of Radiology and Nuclear Medicine, Oslo University Hospital, Oslo, Norway

[2]Department of Physics, Faculty of Mathematics and Natural Sciences, University of Oslo, Oslo, Norway

[3]Department of Physics and Computational Radiology, Division of Radiology and Nuclear Medicine, Oslo University Hospital, Oslo, Norway

[4]The Intervention Centre, Oslo University Hospital, Oslo, Norway

[5]Department of Optometry, Radiography and Lighting Design, University of South-Eastern Norway, Drammen, Norway

[6]Department of Health Sciences Gjøvik, Faculty of Medicine and Health Sciences, NTNU, Gjøvik, Norway

[7]Department of Neurosurgery, Oslo University Hospital, Oslo, Norway

[8]Department of Radiology, Vestfold Hospital Trust, Tønsberg, Norway

[9]Amsterdam UMC, University of Amsterdam, Biomedical Engineering and Physics, Amsterdam, Netherlands

[10]Vilhelm Magnus Laboratory, Department of Neurosurgery, Oslo University Hospital, Oslo, Norway/ Institute for Clinical Medicine, Faculty of Medicine, University of Oslo, Oslo, Norway

**\* Correspondence:**

# Abstract

**Purpose:** To evaluate the quality of deep learning reconstruction for prospectively accelerated intraoperative magnetic resonance imaging (iMRI) during resective brain tumor surgery.

**Materials and Methods:** Accelerated iMRI was performed during brain surgery using dual surface coils positioned around the area of resection. A deep learning (DL) model was trained on the fastMRI neuro dataset to mimic the data from the iMRI protocol. Evaluation was performed on imaging material from 40 patients imaged between 01.11.2021 – 01.06.2023 that underwent iMRI during tumor resection surgery. A comparative analysis was conducted between the conventional compressed sense (CS) method and the trained DL reconstruction method. Blinded evaluation of multiple image quality metrics was performed by two working neuro-radiologists and a working neurosurgeon on a 1 to 5 Likert scale (1=non diagnostic, 2=poor, 3=acceptable, 4=good, 5=excellent), and the favored reconstruction variant.

**Results:** The DL reconstruction was strongly favored or favored over the CS reconstruction for 33/40, 39/40, and 8/40 of cases for reader 1, 2, and 3, respectively. Two of three readers consistently assigned higher ratings for the DL reconstructions, and the DL reconstructions had a higher score than their respective CS counterparts for 72%, 72%, and 14% of the cases for reader 1, 2, and 3, respectively. Still, the DL reconstructions exhibited shortcomings such as a striping artifact and reduced signal.

**Conclusion:** DL shows promise to allow for high-quality reconstructions of intraoperative MRI with equal to or improved perceived spatial resolution, signal-to-noise ratio, diagnostic confidence, diagnostic conspicuity, and spatial resolution compared to compressed sense.

# Introduction

Intraoperative magnetic resonance imaging (iMRI) provides surgeons with updated images to improve maximal safe resection of brain tumors. Evidence indicate that iMRI allows for greater extent of resection without an increase in adverse effects (1–3), and many centers have implemented the technique into clinical practice.

Fast imaging times and high image quality are essential to optimize treatment outcomes and minimize idle time of surgical staff during the iMRI-protocol. The sub-optimal conditions for imaging under which iMRI is performed extend both acquisition time and reduce image quality. Manually placed surface coils prohibit fast parallel imaging through generalized autocalibrating partially parallel acquisitions or sensitivity encoding (4,5) that otherwise accelerates brain MRI. The use of iMRI has increased the duration of surgery from 20 to 60 minutes depending on the imaging protocol and surgical setup (6,7). Additionally, the images are more prone to noise distant from the coil elements, unlike those acquired using conventional head coils. Since its advent, one of the main strides in iMRI has been the improvement of image quality, ranging from low to high field scanner configurations (8). Recent reports highlight the need voiced by surgeons for improved iMR image quality (9).

To address prolonged scan times, compressed sense (CS) (10) may be adopted in iMRI. Still, substantial undersampling leads to image artifacts, poor signal-to-noise ratio, lower resolution, and potentially non-diagnostic or poor-quality images where surgically accessible tumor is not identified. A promising alternative is the utilization of deep learning (DL) for reconstruction, aiming to accelerate MRI and enhance quality through denoising and super-resolution strategies. Key to its success is training DL networks on relevant MRI data that is representative of the foreseen imaging setting (11–14). Initial prospective studies on knee, spine, and brain images have demonstrated that DL networks were able to generalize to anatomy and pathology beyond their training scope (15–18). Further evidence is needed to warrant deploying DL networks in the highly variable iMRI setting.

In this study, we aimed to investigate (1) the generalizability of DL-based reconstruction on prospectively accelerated iMRI during tumor resection surgery after being trained on conventional neuro MRI, and (2) whether DL based reconstruction can generate high resolution iMR images with improved diagnostic quality compared to the currently used protocol and ease the detection of tumor residuals. For this we adopted the densely interconnected residual cascading network (DIRCN) (19) and trained the network on the fastMRI neuro dataset (20,21) with pairs of fully sampled and retrospectively downsampled images, adjusted to match the clinical protocols downsampling scheme. We hypothesized that DL-based reconstruction can successfully generalize to an intraoperative protocol and outperform the current standard of CS method.[1]

# Materials and Methods

## Participants

This study was approved by the Regional Medical Ethics Committee for Oslo University Hospital (REK 367336), and informed consent was acquired prior to surgery according to a broad research approval and biobank (REK 2016/17091). Only patients over 18 years were included.

---

[1] The model weights and training scripts are available at https://github.com/JonOttesen/intraop_recon

This study cohort included 40 patients (mean age 53±14 years; 17 women) with histologically confirmed glioblastoma in 18 patients, 9 astrocytoma (2 grad 2, 4 grade 3, and 3 grade 4), 9 oligodendroglioma (7 grade 2 and 2 grade 3), 1 dysembryoplastic neuroepithelial tumor, 1 brain metastases from a malignant melanoma, one lymphoma, and one tumor with bleeding without confirmed diagnosis due to biopsy quality.

**Acquisition Protocol**

Imaging was performed on a Philips Ingenia 3T MRI using two receiver surface coil elements (Philips Healthcare, Best, The Netherlands). Patients were head clamped and fixed to the operating table. Positioning had to take into consideration the access of the surgeon to the tumor, adequate access to airway management and the comfort of the patient for those who were operated with awake monitoring, in addition to the position within the gantry. Most patients are thus positioned well outside the imaging isocenter. During surgery the tabletop of the surgical table was translocated to the MRI situated in the room next door.

The imaging protocol included Sagittal 3D T1-weighted pre- and post-contrast, and 3D fluid attenuated inversion recovery (FLAIR)-weighted image sets. Additionally, 2D axial turbo spin-echo (T2-weighted) and diffusion MRI were acquired, but not included in the study. The T1-weighted pre- and post-contrast image sets were acquired with a 135-153 x 199 x 384 acquisition matrix with a resolution of 1.25 x 1.24 mm and a slice thickness of 1.74 mm (scan time 3m56s). The FLAIR-weighted images were acquired with a 119-133 x 165 x 336 acquisition matrix with a resolution of 1.49 x 1.50 mm and a slice thickness of 1.50 mm (scan time 5m26s), and two averages. Note, due to scanner settings, certain patient orientations resulted in an extended field-of-view (FOV) oversampling factor of approximately ~2.8 along the posterior anterior direction, with an equal factor of undersampling. In-plane oversampling occurred for 30 of the 40 patients, altering their acquisition matrix to 135-153 x 569 and 119-133 x 165 for the T1-weighted and FLAIR-weighted images, respectively. The field-of-view (FOV) was extended to avoid infolding artifacts from objects outside the FOV.

K-space was sampled with a pseudo-random pattern with acceleration factors of 1.5 and 3.7 for the T1-weighted and FLAIR images respectively. For the images with an oversampling factor, the acceleration factors were increased to 4 and 10 for the T1-weighted and FLAIR-weighted images, respectively.

The on-scanner T1-weighted pre- and post-contrast reconstructions had a zero-filled interpolated reconstruction resolution of 0.78 x 0.78 mm with a slice thickness of 0.87 mm; the FLAIR-weighted scanner reconstructions had a resolution of 0.74 x 0.74 mm with a slice thickness of 0.75 mm. The same scans used for the compressed sense reconstruction were also used for the DL reconstruction.

**Data Processing and Training**

The DIRCN (19) was trained to reconstruct accelerated data and to achieve super-resolution from an in-plane resolution of 1.2-1.6 mm to 0.43-0.9 mm. The model was initialized with random parameters and it was trained on pairs of downsampled and fully sampled scans obtained from the fastMRI neuro dataset (20,21). The dataset consists of 4,469 training and 1,378 validation patients with fully sampled 2D axial brain T1-weighted pre- and post-contrast, FLAIR-weighted, and T2-weighted image sets. The inverse Fourier transform was performed along the z-direction prior to DL reconstruction. The dataset has an in-plane resolution between 0.43 and 0.9 mm, with the median resolution being 0.6875 x 0.6875 mm.

To emulate clinical iMRI protocol settings, the FOV of the data was first randomly cropped in image space to a quadratic or two-fold oversampled image size along the frequency encoding direction. Subsequently, k-space was randomly cropped to ensure an in-plane image resolution between 1.2-1.6 mm - matching the native resolution found in the iMRI protocols used in this study. The downsampled k-space was thereafter masked with a Poisson-disc sampling pattern. This pattern was generated randomly by either SigPy (22) or the BART Toolbox (23) with randomized acceleration factors that matched the acceleration factors seen in the clinical protocol, i.e., a 1.5-, 3.7-, 4-, and 10-times acceleration factors. By utilizing two sampling pattern methods, we aimed to better encompass the proprietary vendor sampling masks. The model was trained to reconstruct 2D axial slices.

The Adam optimizer (24) was used with a cosine annealing learning rate scheduler (25). The model was trained for a total of 200 epochs with a learning rate of $lr = 1e-3$ to minimize the structural similarity index measure (26), and one epoch iterated over 25,000 training examples. Model evaluation was performed after each epoch, and the epoch with the lowest validation loss was selected for inference. Training was performed with a batch size of two, utilizing two NVIDIA RTX3090 graphical processing units (GPUs).

**Reconstruction and Image Processing**

The intraoperative scans were zero-padded in k-space to reach the desired target resolution, reconstructed by the DIRCN model, and bias field corrected (27). Repeated samples were averaged in k-space before DL reconstruction.

In preparation for the quality assessment, the on-scanner images were up-sampled such that their resolution matched the DIRCN-reconstructions. All images were co-registered to the on-scanner T1 pre-contrast image (27) and de-faced before evaluation (28).

An example of the raw protocol images, the training regime, and reconstruction process is illustrated in Figure 1.

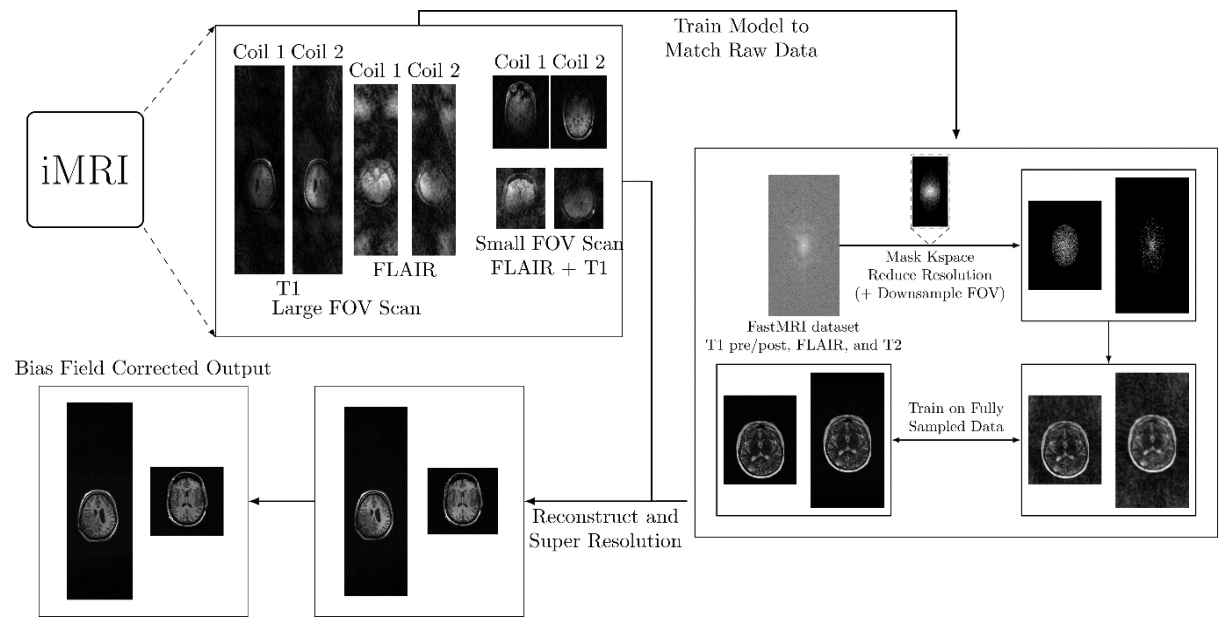

*Figure 1: An Illustration of the raw intraoperative magnetic resonance imaging (iMRI) data before any reconstruction, the model training/reconstruction regime, and the resulting deep learning (DL) reconstruction with and without bias field correction. The iMRI scans have two different imaging protocols with a large field-of-view (FOV) or a small FOV. A DL model was trained on the fastMRI dataset to match the iMRI protocol with respect to the masked k-space, FOVs, and resolution. The model was used to reconstruct the prospective iMRI scans followed by bias field correction.*

**Qualitative Assessment**

Prior to quality assessment, the imaging variants (DL and CS) were randomly assigned with a pseudo random number generator in Python to "A" or "B" for each patient to ensure blinded assessment. Assessment of "A" and "B" for a given patient was performed at the same time or in a similar timeframe.

Image quality assessment was performed by two neuroradiologists, with 10 and 19 years of experience and a neurosurgeon. The readers were provided with data in DICOM or NIfTI format, and they had the freedom to choose their preferred reading software. The images were ranked on a Likert scale: 1=non diagnostic, 2=poor, 3=acceptable, 4=good, and 5=excellent on the following metrics: imaging artifacts, perceived spatial resolution, anatomic conspicuity, diagnostic confidence, signal-to-noise (SNR), and contrast. In addition, the CS and DL were blindly ranked based on preference with 1=strongly favors A, 2=favors A, 3=indifferent, 4=favors B, and 5=strongly favors B. A two-sample Wilcoxon signed rank test was performed to test the statistical significance of the assessment. P-values are reported in ranges from < 0.05 (0.01 – 0.05), < 0.01 (0.001 – 0.01), and < 0.001.

Given the iMRI setup, with the dual surface coil setup, we instructed the readers to solely focus on the peritumoral area, in line with the focus of the intraoperative setting. In addition, readers gave their overall preference based on all combined data per patient.

## Results

The mean and standard deviation for the qualitative assessments of the DL and CS image variants from the three expert readers are given in Table 1. There was a significant difference ($p<0.05$) in the scores between DL and CS reconstructed image for all metrics/readers, except for two metrics from reader 3. Across the six-evaluation metrics, DL achieved significantly higher scores than CS for both reader 1 and 2, except for "image artifacts" by reader 1. For reader 3, the CS scored significantly better than the DL counterpart for all metrics except contrast and perceived spatial resolution.

Averaged across all metrics, the DL reconstructions were rated with a higher score than CS reconstruction for 72%, 72%, and 14% of the ratings, for reader 1, 2, and 3, respectively. Conversely, the CS reconstruction was preferred over DL reconstruction for 8%, 7%, and 37% of the ratings, respectively for reader 1, 2, and 3. The violin plot distribution of the qualitative assessments is shown in Figure 2. Of note, since the assessment criteria were discrete, no interpolation between the metrics was performed.

The distribution of the preferred reconstruction is given in Figure 3. The DL reconstruction was favored or strongly favored in 33/40, 39/40, and 8/40 of the cases for reader 1, 2, and 3, respectively. The CS reconstructions were favored or strongly favored in 1/40, 1/40, and 18/40 of the cases for reader 1, 2, and 3, respectively.

Representative images of the DL and CS reconstructions are displayed in Figure 4. We note that the DL reconstructions show a higher level of detail in the resection area, most pronounced in the FLAIR-weighted images. In some cases, imaging artifacts were reported by the readers and DL specific artifacts are shown in Figure 5. In particular, the most common artifacts mentioned were high noise/grainy images and reduced signal distant from the receiver coils. Upon later inspection by the authors, we noted that the high noise artifact can be classified into two categories: whole brain noise and slice specific high noise levels.

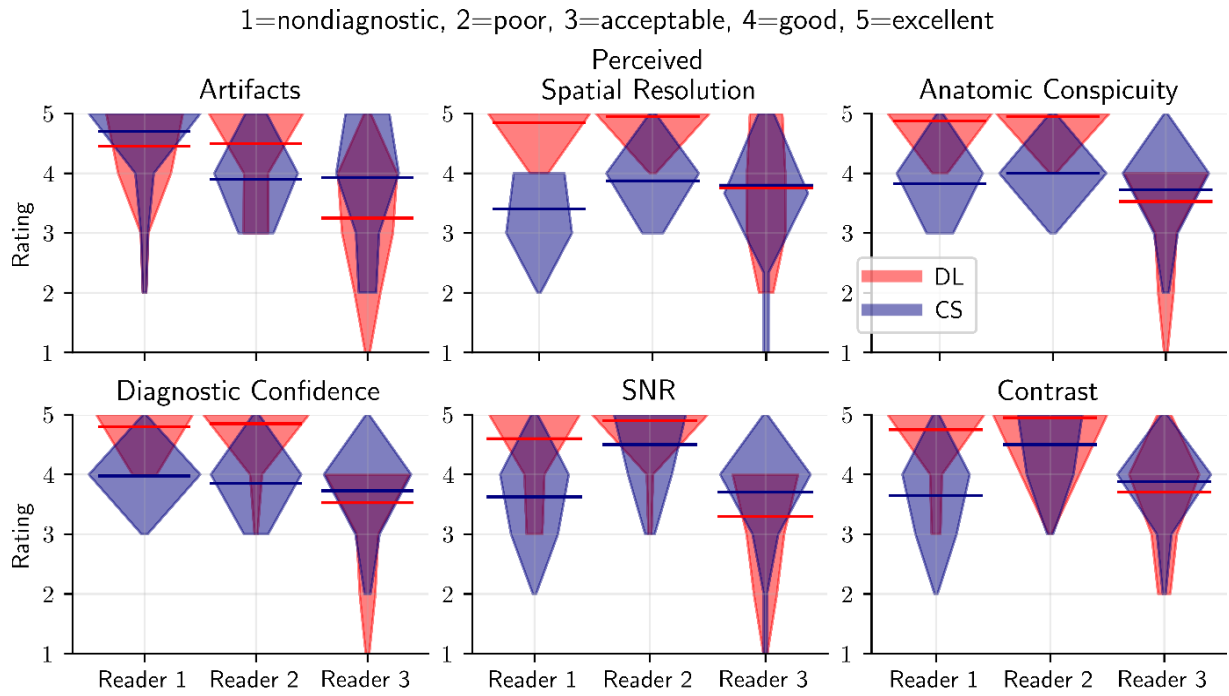

*Figure 2: A violin plot of the qualitative assessment of the deep learning (DL) and compressed sense (CS) reconstructions for 40 intraoperative patients from three expert readers. The mean of the DL and CS assessments are shown with the colored lines. No smoothing has been used in the violin plot due to the discrete nature of the assessment.*

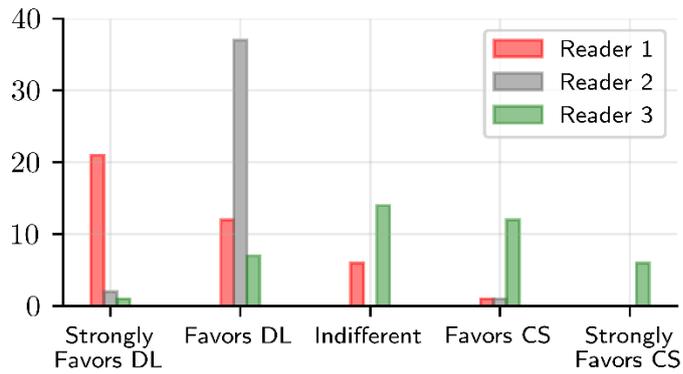

*Figure 3: The preferred reconstruction variant (Deep Learning – DL, Compressed Sense – CS) among the three expert readers.*

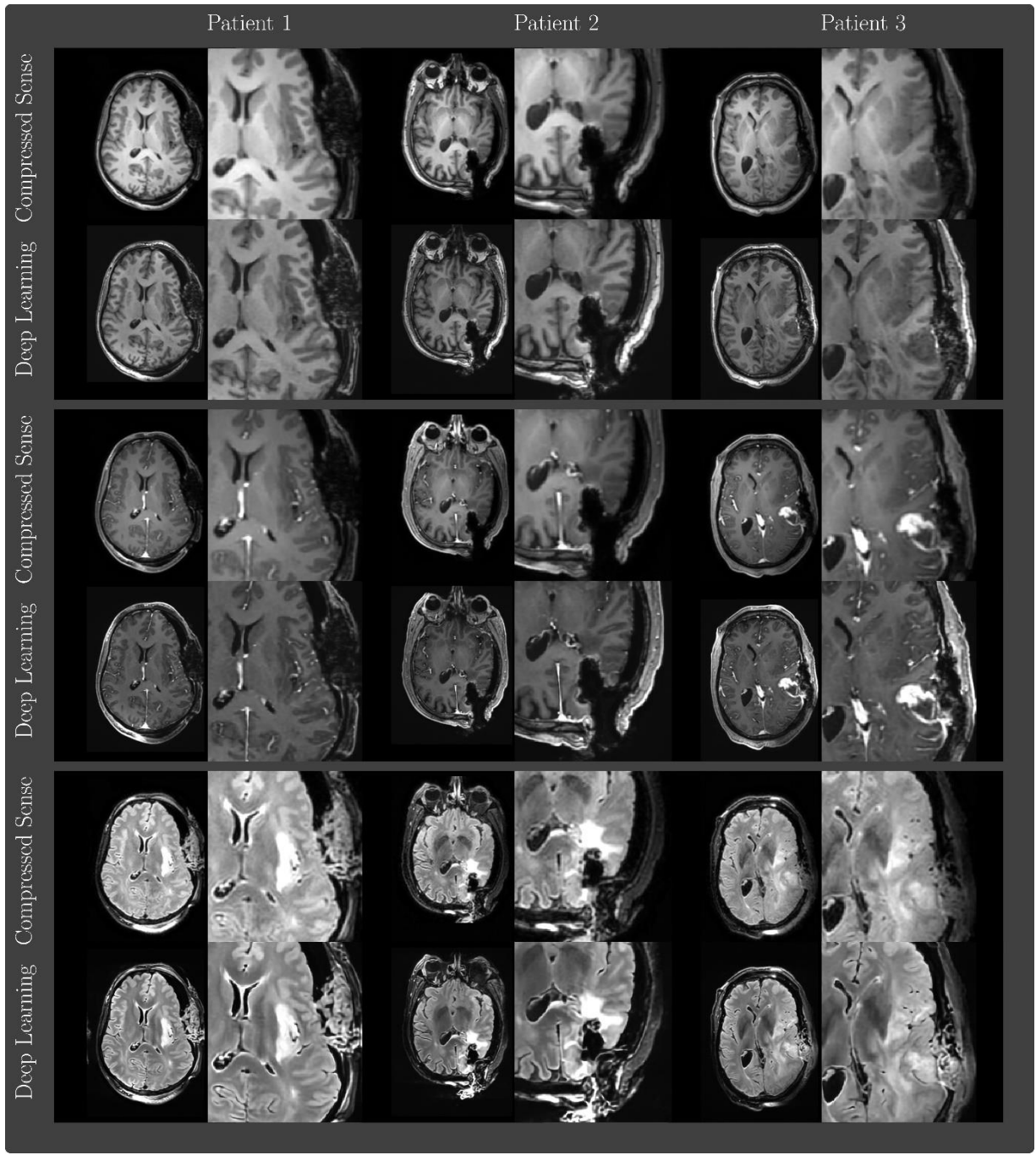

*Figure 4: Representative examples of the deep learning (DL) and compressed sense (CS) reconstructions from four different patients, depicting T1-weighted pre-contrast, T1-weighted post-contrast, and FLAIR-weighted scans. Window leveling was chosen between 0.05 and 0.995 percentiles.*

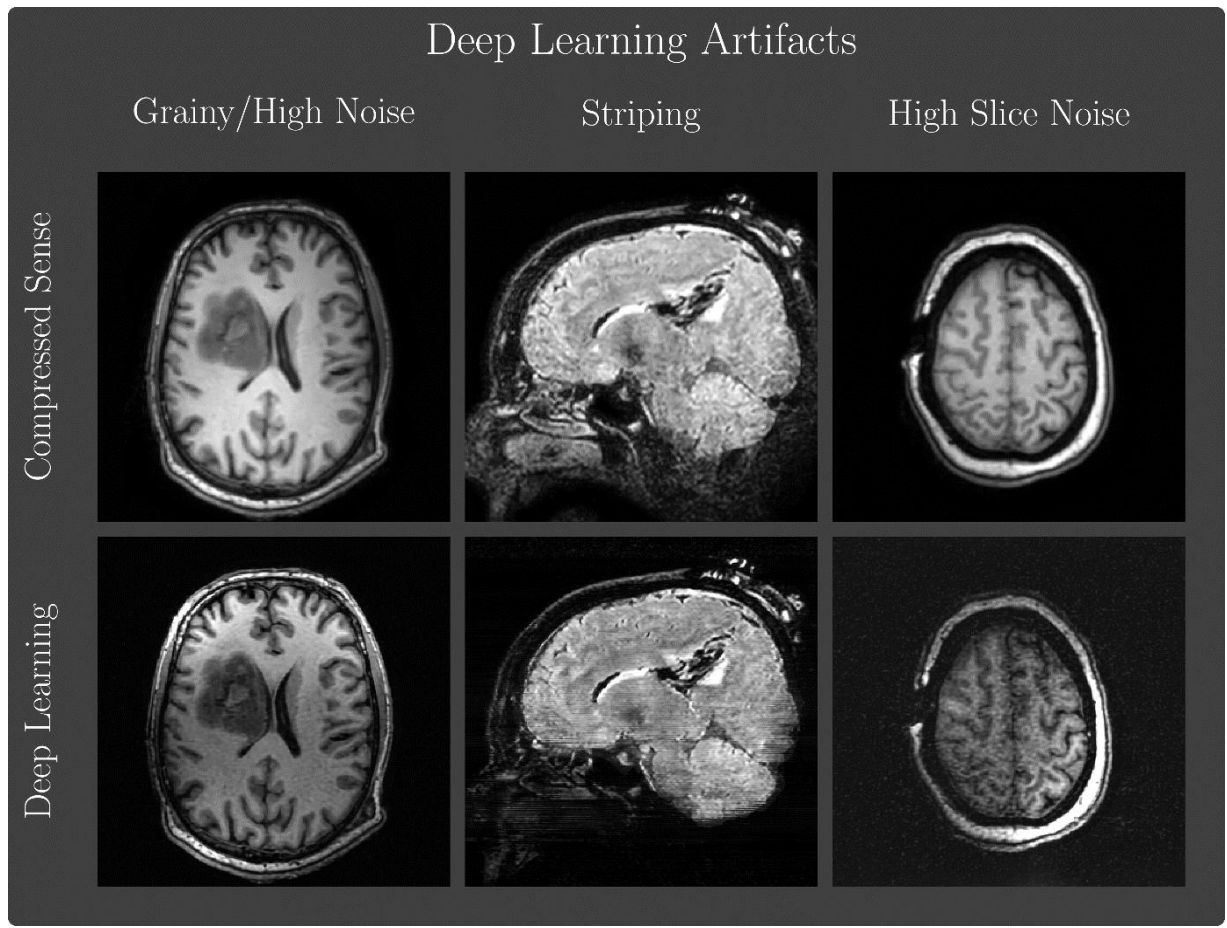

*Figure 5: Examples of deep learning (DL) specific artifacts. This includes high noise levels for a given scan, striping-like artifacts, and high noise for one slice in an image.*

## Discussion

In this study, we have adapted and trained a DL reconstruction network to reconstruct prospectively undersampled iMRI data. The model was qualitatively evaluated and compared to the CS reconstructions by three expert readers on a cohort of 40 patients who underwent iMRI. DL was found to reconstruct iMRI with significantly better image quality compared to CS for two of the three readers. Notably, the neuroradiologists (readers 1 and 2) favored overall DL over CS to a larger extent than the neurosurgeon (reader 3). Reader 3 assessed that the CS reconstruction had improved image quality compared to the DL equivalent. The difference in image quality assessment reflects the subjective nature of the interpretation. Further, iMRI is different from conventional MRI diagnostics in that radiologists rarely review these images whereas they are a central part of the neurosurgeon's intraoperative assessment of the extent of tumor resection. In discussing the results with the expert reviewers, it was noted that the focus of the neurosurgeon was mainly on the extent of resection, whereas the radiologist focused more on image quality and diagnostic quality in the area-of-interest.

iMRI requires specialized operating and scanning equipment (8), relying on the use of flexible surface coils to adjust to the required head position dictated by the tumor location. The required coil setup typically results in low SNR in areas distant from the target tumor region. This is partially compensated for by the coil sensitivity maps used in the CS reconstruction, but not the DL variants. As such, the low SNR artifact was particularly visible in the DL images, as pointed out by reader 3. The low intensity coupled with the 2D nature of the reconstruction gave rise to pronounced striping artifacts in some cases with DL reconstruction. Conversely, the CS reconstructions suffered more from low SNR artifacts. It is worth noting that most of these artifacts occurred outside the resection area of interest due to the coil placement. Despite the artifacts, the DL model showcased high quality reconstruction and super-resolution, and all reviewers graded all DL reconstructed images were of good or acceptable quality.

Importantly, DL-specific artifacts were rare, but in some cases worrisome when brain regions close to the area of resection had low signal as can be seen in Figure 6 that highlights three cases with worrisome DL-related artifacts. Note, the artifact in the first case stems from the bias-field-correction. The two remaining cases had reduced signal in the area-of-resection in one or more of the scans taken. Upon close inspection of the raw non-reconstructed images, it was seen that the affected scans either had a high degree of noise or very low signal in affected area. Although the artifacts were a rare occurrence, it is important to be addressed as it can impact surgical decision making.

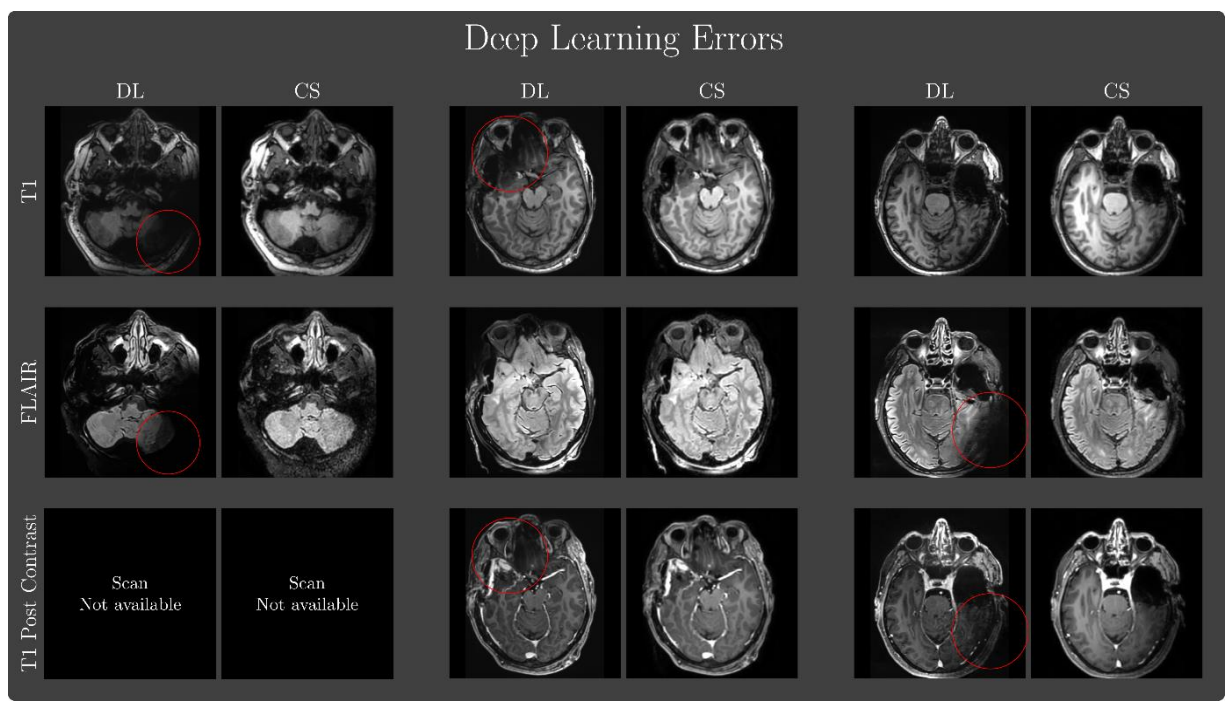

*Figure 6: Three cases highlighted by the neurosurgeon where the DL reconstructions had considerable artifacts. For the two last patients, the artifacts affected the area of resection for one or more of the scans. The first patient, i.e., columns 1 and 2 did not have a T1-weighed post-contrast series, and the cell was therefore left blank.*

The majority of works in DL-based reconstruction of MRI has focused on retrospectively downsampled data. Recent studies have demonstrated that DL can successfully be adapted on prospectively downsampled data, particularly in contexts with relatively normal pathology and scanner configurations (15,17,18,29). This study aims to validate the generalizability of DL reconstruction of highly non-conventional brain images.

Except for imaging artifacts, DL reconstruction significantly outperformed the CS reconstruction on all assessment metrics by both reader 1 and 2. In addition, since the fastMRI dataset was imaged exclusively on Simens scanners (20,21), this study implies that DL models exhibit generalizability across vendors and scanner setups even on an unordinary scanner setup that is iMRI.

The findings in this study should be interpreted with multiple limitations in mind. First, the study included a relatively limited patient cohort size of 40 patients from a single site.

Additional patients from multiple sites would need to further validate the generalizability of the model. Second, the DL reconstructions were only compared to their respective CS reconstruction, and not the ground truth in a quantitative analysis. Acquiring fully sampled ground truth images in an iMRI setting would be very time consuming and therefore not feasible for ethical reasons and the clinically used CS based accelerated reconstruction was therefore considered the best standard of comparison. Third, the number of MRI sequences per patient varied, so each scoring is not for the same number of sequences. We opted for this method of assessment since an expert reader would evaluate the case "as a whole" with all scans. Fourth, although the readers were blinded, and the DL/CS scans were randomized per patient; it is feasible to accurately guess which is which by an expert reader due to the general imaging characteristics of the different reconstruction methods.

Future work could explore the effect of further increase in acceleration factor. Here, we opted against retrospectively downsampling the prospective data to simulate further acceleration as we wanted the study to solely focus on prospectively undersampled data. To properly evaluate the potential gains from the improved image quality, a "during surgery" evaluation would strengthen the findings in this study. Additional focus should be placed on evaluation of the extent of resection.

## Conclusion

DL-based MR reconstruction allows for high quality reconstructions of prospectively undersampled intraoperative MRI. Two of the expert readers favored the DL reconstruction over the CS counterpart, and one reader favored the CS reconstruction. Still, further work is needed to account for low coil sensitivity distant from the area of resection and alleviate the striping artifact seen from the 2D nature of the reconstruction.

## Acknowledgments

The project was supported by the Norwegian South-Eastern Health Authority (grant numbers 2021031, 2016102, 2017073 and 2013069). The Research Council of Norway grant numbers 261984, 325971. The Norwegian Cancer Society grant numbers 6817564 and 3434180. The European Research Council grant number 758657-ImPRESS.

## Conflict of Interest

M.W.A. Caan is shareholder of Nico-lab International Ltd.

*Table 1: The mean and standard deviation for the qualitative assessment of image artifacts, perceived spatial resolution, anatomic conspicuity, diagnostic confidence, signal-to-noise (SNR), and contrast from three expert readers on deep learning (DL) and compressed sense (CS) reconstructed images. Due to the coil setup, the area of assessment was the area of resection and peritumoral. P-values are reported in ranges from < 0.05 (0.01 – 0.05), < 0.01 (0.001 – 0.01), and < 0.001.*

| Feature | Reader 1 | | | Reader 2 | | | Reader 3 | | |
|---|---|---|---|---|---|---|---|---|---|
| | DL | CS | P-Value | DL | CS | P-Value | DL | CS | P-Value |
| Image Artifacts | 4.5±0.7 | 4.7±0.7 | < 0.05 | 4.5±0.8 | 3.9±0.6 | < 0.01 | 3.3±0.9 | 3.9±1.0 | < 0.05 |
| Perceived Spatial Resolution | 4.9±0.4 | 3.4±0.5 | < 0.001 | 4.95±0.2 | 3.9±0.5 | < 0.001 | 3.8±1.0 | 3.8±0.7 | 0.75 |
| Anatomic Conspicuity | 4.9±0.3 | 3.8±0.5 | < 0.001 | 4.95±0.2 | 4.0±0.3 | < 0.001 | 3.5±0.8 | 3.7±0.6 | < 0.05 |
| Diagnostic Confidence | 4.8±0.4 | 4.0±0.3 | < 0.001 | 4.9±0.4 | 3.9±0.5 | < 0.001 | 3.5±0.8 | 3.7±0.6 | < 0.05 |
| Signal-to-Noise (SNR) | 4.6±0.7 | 3.6±0.6 | < 0.001 | 4.9±0.4 | 4.5±0.6 | < 0.01 | 3.3±0.8 | 3.7±0.7 | < 0.01 |
| Contrast | 4.8±0.6 | 3.7±0.6 | < 0.001 | 4.95±0.3 | 4.5±0.6 | < 0.001 | 3.7±0.8 | 3.9±0.5 | 0.17 |